\documentclass[11pt]{article}
\usepackage{fullpage}
\usepackage{cite}
\usepackage{graphicx}
\usepackage{amsmath}
\usepackage{amssymb}
\title{}
\date{}
\renewcommand{\vec}[1]{\mbox{\boldmath$ #1 $}}

\def\beq{\begin{equation}}
\def\eeq{\end{equation}}
\allowdisplaybreaks
\begin{document}
\bibliographystyle{utphys}
\newcommand{\msbar}{\ensuremath{\overline{\text{MS}}}}
\newcommand{\DIS}{\ensuremath{\text{DIS}}}
\newcommand{\abar}{\ensuremath{\bar{\alpha}_S}}
\newcommand{\bb}{\ensuremath{\bar{\beta}_0}}
\newcommand{\rc}{\ensuremath{r_{\text{cut}}}}
\newcommand{\Nd}{\ensuremath{N_{\text{d.o.f.}}}}
\setlength{\parindent}{0pt}

\titlepage
\begin{flushright}
QMUL-PH-16-18
\end{flushright}

\vspace*{0.5cm}

\begin{center}
{\bf \Large Exact solutions for the biadjoint scalar field}

\vspace*{1cm}
\textsc{C. D. White\footnote{Christopher.White@glasgow.ac.uk} } \\

\vspace*{0.5cm} School of Physics and Astronomy, University of Glasgow,\\ Glasgow G12 8QQ, Scotland, UK\\

\vspace*{0.5cm} Centre for Research in String Theory, School of
Physics and Astronomy, \\
Queen Mary University of London, 327 Mile End
Road, London E1 4NS, UK\\

\end{center}

\vspace*{0.5cm}

\begin{abstract}
Biadjoint scalar theories are novel field theories that arise in the
study of non-abelian gauge and gravity amplitudes. In this short
paper, we present exact nonperturbative solutions of the field
equations, and compare their properties with monopole-like solutions
in non-abelian gauge theory. Our results may pave the way for
nonperturbative studies of the double copy.
\end{abstract}

\vspace*{0.5cm}

\section{Introduction}
\label{sec:intro}

There has recently been much attention on the relationship between
different types of field theory. A notable example is the {\it double
  copy} of refs.~\cite{Bern:2008qj,Bern:2010ue,Bern:2010yg}, which
relates perturbative scattering amplitudes in non-abelian gauge and
gravity theories. At tree-level this has a string theoretic
explanation from the well-known KLT
relations~\cite{Kawai:1985xq}. However, the double copy remains a
conjecture at loop level, where it has been tested up to four
loops~\cite{Bern:2010ue,Bern:1998ug,Green:1982sw,Bern:1997nh,Oxburgh:2012zr,Carrasco:2011mn,Carrasco:2012ca,Mafra:2012kh,Boels:2013bi,Bjerrum-Bohr:2013iza,Bern:2013yya,Bern:2013qca,Nohle:2013bfa,Bern:2013uka,Naculich:2013xa,Du:2014uua,Mafra:2014gja,Bern:2014sna,Mafra:2015mja,He:2015wgf,Bern:2015ooa,Mogull:2015adi,Chiodaroli:2015rdg},
to all orders in certain kinematic
limits~\cite{Oxburgh:2012zr,Saotome:2012vy,Vera:2012ds,Johansson:2013nsa,Johansson:2013aca,Melville:2013qca},
and in the self-dual sector~\cite{Monteiro:2011pc}. A related
programme of work relating amplitudes in different theories has arisen
from the {\it CHY equations} of
refs.~\cite{Cachazo:2013iea,Cachazo:2013hca,Cachazo:2013gna,Cachazo:2013iaa}. These
express tree-level amplitudes in gauge and gravity theories in terms
of an abstract integral representation involving punctures on the
Riemann sphere, which is itself reminiscent of string theory: indeed,
the CHY equations can be obtained from ambitwistor string
theory~\cite{Mason:2013sva,Geyer:2014fka,Casali:2015vta}, which
framework has recently been used to extend the former to one-loop
level~\cite{Geyer:2015bja,Geyer:2015jch}.\\

All of the above work focuses on perturbative properties of the
respective field theories. If the double copy is truly correct, it
must also be possible to relate exact classical solutions of gauge
theory and gravity, where these are known. First steps in this
direction were undertaken in ref.~\cite{Monteiro:2014cda}, which found
an infinite class of classical solutions in General Relativity (namely
{\it stationary Kerr-Schild metrics}), possessing well-defined
counterparts in a gauge theory. Examples include the Schwarzschild and
Kerr black holes, which give rise to pointlike and rotating
distributions of charge in the gauge theory respectively. One may
generalise the former case to a pointlike dyon in the gauge theory,
which double copies to the Taub-NUT solution in
GR~\cite{Luna:2015paa}. More recently, ref.~\cite{Luna:2016due}
examined an arbitrarily accelerating (and radiating) particle in gauge
theory and gravity, and further work has examined whether the source
terms in both theories are physical~\cite{Ridgway:2015fdl}. \\

Both the double copy and the CHY equations relate solutions in
well-known gauge theories and gravity. This is not the whole story,
however, in that one also finds more exotic theories whose amplitudes
are related to gauge theory. In this paper we focus on {\it biadjoint
  scalar theories}, containing a field ${\bf \Phi}=\Phi^{aa'}{\bf
  T}^a\,\tilde{\bf T}^{a'}$, where $\Phi^{aa'}\in{\mathbb R}$, and
$\{{\bf T}^a\}$ and $\{\tilde{\bf T}^a\}$ are generators of two
(possibly different) Lie algebras:
\begin{equation}
[{\bf T}^{a},{\bf T}^b]=if^{abc}{\bf T}^c,\quad
[\tilde{\bf T}^{a},\tilde{\bf T}^{b}]
=i\tilde{f}^{abc}\tilde{\bf T}^{c}.
\label{Lie}
\end{equation}
We write the Lagrangian defining the theory as
\begin{equation}
{\cal L}=\frac12\partial^\mu\Phi^{aa'}\partial_\mu\Phi^{aa'}+\frac{y}{3}
f^{abc}\tilde{f}^{a'b'c'}\Phi^{aa'}\Phi^{bb'}\Phi^{cc'},
\label{Lagrangian}
\end{equation}
which gives rise to the equation of motion
\begin{equation}
\partial^2\Phi^{aa'}-yf^{abc}\tilde{f}^{a'b'c'}\Phi^{bb'}\Phi^{cc'}=0.
\label{EOM}
\end{equation}
This is a novel theory, which we may think of as describing a scalar
field with two types of charge. Although such theories seem not to be
directly applicable to nature, they nevertheless have an importance of
their own. Firstly, they underlie the structure of gauge and gravity
theories, in both the double copy and CHY approaches. For example,
ref.~\cite{BjerrumBohr:2012mg} explains how to use biadjoint theories
as building blocks for constructing amplitude numerators in gauge
theory, in such a way as to make the double copy manifest. Secondly,
the field equation of eq.~(\ref{EOM}) was crucial in
ref.~\cite{Monteiro:2014cda} for arguing that the classical double
copy considered there was related to the BCJ story for perturbative
amplitudes. Finally, scalar theories are often simpler than gauge or
gravity theories. Thus, one may potentially use exact solutions of the
former to investigate solutions of the latter. This is essentially the
spirit of the classical double copy of
refs.~\cite{Monteiro:2014cda,Luna:2015paa,Luna:2016due,Ridgway:2015fdl}.
However, the solutions considered there were very special in that both
the Yang-Mills and Einstein equations linearised. Correspondingly, the
cubic term in the Lagrangian of eq.~(\ref{Lagrangian}) vanished.\\

The aim of this paper is to derive exact solutions of eq.~(\ref{EOM})
in which the interaction term is nonzero. We will focus on static
solutions, and furthermore those that are fully nonperturbative, in
that they involve inverse powers of the coupling constant $y$ in
eqs.~(\ref{Lagrangian}, \ref{EOM}). To the best of our knowledge, this
has not been previously carried out. We will study a number of
nontrivial solutions, and contrast these with nonperturbative
solutions in non-abelian gauge theory. Whilst a full understanding of
any double-copy like property is beyond the scope of this paper, we
will see intriguing hints that nonperturbative solutions in biadjoint
theory and gauge theory are related. Thus, our results constitute an
important step in being able to probe nonperturbative aspects of the
double copy and / or CHY frameworks.\\

The structure of the paper is as follows. In section~\ref{sec:np} we
consider a number of simple ans\"{a}tze, and a first non-perturbative
solution, applicable when the two Lie algebras coincide with each
other. In section~\ref{sec:SU2} we focus specifically on the case in
which both Lie algebras are SU(2), finding a more general form for the
scalar field. Throughout, we compare our solutions with similar
solutions in non-abelian gauge theory. Finally, we discuss our results
and conclude in section~\ref{sec:conclude}.

\section{A first nonperturbative solution}
\label{sec:np}

As a first attempt at solving eq.~(\ref{EOM}), one may try the form
\begin{equation}
\Phi^{aa'}(x)=\chi^{a}(x)\xi^{a'}(x),
\label{factorised}
\end{equation}
in which each adjoint index $a$, $a'$ has a nontrivial kinematic
dependence associated with it, but the two types of charge do not talk
to each other. Substituting this into eq.~(\ref{EOM}), one immediately
finds (from the antisymmetry of the structure constants) that the
interaction term vanishes, leaving
\begin{equation}
\partial^2(\chi^{a}(x)\xi^{a'}(x))=0.
\label{noninteract}
\end{equation}
One may further classify general solutions of this form but, being
solutions of the free theory, they are not interesting for our present
purpose, which is to find nonperturbative solutions of the full field
equation. However, it is already interesting that, in order to find a
nonperturbative solution, the two types of charge in the biadjoint
theory must be inextricably linked. \\

Let us now restrict ourselves to the case where both sets of structure
constants in eq.~(\ref{EOM}) come from the same Lie algebra. One may
then write the ansatz
\begin{equation}
\Phi^{aa'}=\delta^{aa'}S(r),
\label{ansatz1}
\end{equation}
where we assume spherical symmetry, and $r$ is the radial
coordinate. Substituting eq.~(\ref{ansatz1}) into eq.~(\ref{EOM})
yields~\footnote{Note that we use the spacetime metric $(+,-,-,-)$
  throughout.}
\begin{equation}
\delta^{aa'}\nabla^2 S(r)+yf^{abc}f^{a'bc}S^2(r)=0.
\label{Seq1}
\end{equation}
In the second term, we may use the colour algebra
\begin{align}
f^{abc}f^{a'bc}&={\rm Tr}\left[{\bf T}^a {\bf T}^{a'}\right]\notag\\
&=\delta^{aa'}T_A
\label{col}
\end{align}
where ${\bf T}^a$ is a colour generator in the adjoint representation,
and $T_A$ the relevant normalisation constant. Then eq.~(\ref{Seq1})
becomes
\begin{equation}
\delta^{aa'}\left[\frac{1}{r^2}\frac{d}{dr}\left(
r^2\frac{d S(r)}{dr}\right)+y\, T_A\, S^2(r)\right]=0,
\label{Seq2}
\end{equation}
where we have used the usual form of the Laplacian in spherical
polars. At this point one may scale 
\begin{equation}
S(r)=\frac{\bar{S}(r)}{y\,T_A}\quad\Rightarrow\quad
\frac{1}{r^2}\frac{d}{dr}\left(r^2\frac{d\bar{S}(r)}{dr}\right)
+\bar{S}^2(r)=0,
\label{Seq3}
\end{equation}
and look for a solution of the form $\bar{S}(r)=Kr^\alpha$, which
finally leads to the nontrivial field
\begin{equation}
\Phi^{aa'}=-\frac{2\delta^{aa'}}{y\,T_A\,r^2}.
\label{phisol1}
\end{equation}
As already mentioned in the introduction, this is a fully
nonperturbative solution, in that it contains an inverse factor of the
coupling constant. It falls off rapidly as $r\rightarrow\infty$, and
is singular at $r=0$. Thus, it represents a point-like excitation
localised at the origin. To further physically interpret this
solution, we may study its energy. From eq.~(\ref{Lagrangian}), the
Hamiltonian density is given by
\begin{align}
{\cal H}&=\frac12\left[(\dot{\Phi}^{aa'})^2+\nabla\Phi^{aa'}\cdot 
\nabla\Phi^{aa'}\right]-\frac{y}{3}f^{abc}\tilde{f}^{a'b'c'}
\Phi^{aa'}\Phi^{bb'}\Phi^{cc'}\notag\\
&=\frac{32}{3}\frac{\cal N}{y^2 T_A^2}\frac{1}{r^6},
\label{Hamiltonian}
\end{align}
where in the second line we have substituted eq.~(\ref{phisol1}), and
defined ${\cal N}$ to be the dimension of the (common) Lie group. The
energy can be obtained by imposing a short-distance radial cutoff, to
get
\begin{align}
E&=\int d^3\vec{x}\,{\cal H}\notag\\
&=\frac{128\pi}{3}\frac{{\cal N}}{y^2 T_A^2}
\int_{r_0}^\infty \frac{dr}{r^4}\notag\\
&=\frac{128\pi}{9}\frac{\cal N}{y^2 T_A^2}\frac{1}{r_0^3}.
\label{E}
\end{align}
This is divergent at small distances, analogous to the case of a
point-like charge in gauge theory, such as the singular solutions of
ref.~\cite{Rosen:1972uu}. However, at large distances the energy is
bounded, leading to a well-defined result upon assuming a finite
charge radius. \\

In this section we have seen an exact, non-perturbative solution to
the biadjoint field equation, valid when both Lie groups are the same
as each other. A more interesting solution is possible if this common
group is taken to be SU(2), as we explore in the following section.

\section{Further solutions for $SU(2)\otimes SU(2)$}
\label{sec:SU2}

In this section we choose both sets of structure constants in
eqs.~(\ref{Lagrangian}, \ref{EOM}) to correspond to SU(2), so that the
field equation becomes
\begin{equation}
\nabla^2\Phi^{aa'}+y\epsilon^{abc}\epsilon^{a'b'c'}
\Phi^{bb'}\Phi^{cc'}=0,
\label{EOM2}
\end{equation}
where $\epsilon^{abc}$ is the Levi-Cevita symbol, and we have again
focused on static solutions. It is then possible to write the
following ansatz for the field:
\begin{equation}
\Phi^{aa'}=A(r)\delta^{aa'}+B(r)x^a\,x^{a'}+C(r)\epsilon^{aa'd}x^d.
\label{Phiansatz}
\end{equation}
Note that this ansatz contains mixing between spacetime and colour
indices, analogous to nonperturbative solutions in non-abelian gauge
theory~\cite{Prasad:1975kr,Bogomolny:1975de,Julia:1975ff,'tHooft:1974qc,Polyakov:1974wq}. This
requirement motivates our choice of SU(2) as the common Lie
group. Substituting eq.~(\ref{Phiansatz}) into eq.~(\ref{EOM2}), one
obtains
\begin{align}
&\delta^{aa'}\left[A''+\frac{2A'}{r}+2B+2yA(A+Br^2)\right]
+x^{a}x^{a'}\left[B''+\frac{6B'}{r}+2y(C^2-AB)\right]\notag\\
&\quad+\epsilon^{aa' d}x^d\left[C''+\frac{4C'}{r}+2yC(A+Br^2)\right]=0.
\label{Phisol1}
\end{align}
Linear independence of each term then results in the three coupled
nonlinear equations
\begin{align}
A''+\frac{2A'}{r}+2B+2yA(A+Br^2)&=0\label{A};\\
B''+\frac{6B'}{r}+2y(C^2-AB)&=0\label{B};\\
C''+\frac{4C'}{r}+2yC(A+Br^2)&=0.\label{C}
\end{align}
As in the previous section, one may scale out the coupling $y$ by defining $\bar{A}=A/y$ etc., so that eqs.~(\ref{A}--\ref{C}) become
\begin{align}
\bar{A}''+\frac{2\bar{A}'}{r}+2\bar{B}+2\bar{A}(\bar{A}
+\bar{B}r^2)&=0;\label{A2}\\
\bar{B}''+\frac{6\bar{B}'}{r}+2(\bar{C}^2-\bar{A}\bar{B})&=0;\label{B2}\\
\bar{C}''+\frac{4\bar{C}'}{r}+2\bar{C}(\bar{A}+\bar{B}r^2)&=0.\label{C2}
\end{align}
One may next try a power-like solution of the form
\begin{equation}
\bar{A}=ar^\alpha,\quad \bar{B}=b r^\beta,\quad \bar{C}=cr^\gamma.
\label{ABCsol}
\end{equation}
Upon substituting into eqs.~(\ref{A2}--\ref{C2}), one finds
\begin{equation}
\alpha=-2,\quad \beta=-4,\quad \gamma=-3,
\label{alphavals}
\end{equation}
together with the coupled equations
\begin{align}
(a+1)(a+b)=-4b+2c^2-2ab=c(a+b)=0.
\label{abcsol}
\end{align}
A trivial solution is $a=-1$, $b=c=0$, which reproduces
eq.~(\ref{phisol1}) for the special case of SU(2) (note that $T_A=2$
in that case). In addition, there is the one-parameter family of
solutions
\begin{equation}
b=-a=k,\quad c=\pm\sqrt{2k-k^2},
\label{abcsol2}
\end{equation}
where $0\leq k\leq 2$ if $\Phi^{aa'}\in\mathbb{R}$. The complete field is
then given by
\begin{equation}
\Phi^{aa'}=\frac{1}{yr^2}\left[-k\left(\delta^{aa'}-\frac{x^a\,x^{a'}}
{r^2}\right)
\pm\sqrt{2k-k^2}\,\frac{\epsilon^{aa'd}x^d}{r}\right].
\label{phisol2}
\end{equation}
Using the Hamiltonian of eq.~(\ref{Hamiltonian}), the energy of this
solution is given by
\begin{equation}
E=\frac{16\pi k}{y^2r_0^3},
\label{Esol2}
\end{equation}
where again $r_0$ is a small-distance cutoff. As for the solution of
eq.~(\ref{phisol1}), this represents a pointlike object at the
origin. \\

The two solutions of eqs.~(\ref{phisol1}, \ref{phisol2}) have
different boundary conditions, namely that of spherical and
cylindrical symmetry respectively. We may then ask whether
eq.~(\ref{ABCsol}) is the most general solution satisfying cylindrical
symmetry, by instead looking for a solution of the form of
eq.~(\ref{Phiansatz}), but imposing only the requirement
\begin{equation}
B(r)=-\frac{A(r)}{r^2}\quad \Rightarrow \quad \bar{B}(r)=
-\frac{\bar{A}(r)}{r^2}.
\label{symcond}
\end{equation}
A similar ansatz was used to find cylindrically symmetric
multi-instanton solution of the Yang-Mills equations in
ref.~\cite{Witten:1976ck}, as well as the extended Yang-Mills
solutions of
refs.~\cite{Prasad:1975kr,Bogomolny:1975de,Singleton:1999ty}. In the
present case, substituting eq.~(\ref{symcond}) into
eq.~(\ref{A2}--\ref{C2}) yields
\begin{align}
\bar{A}''+2\frac{\bar{A}'}{r}-\frac{2\bar{A}}{r^2}&=0;\label{A3}\\
-\frac{\bar{A}''}{r^2}-\frac{2\bar{A}'}{r^3}+\frac{6\bar{A}}{r^4}
+2\bar{C}^2+\frac{2\bar{A}^2}{r^2}&=0;\label{B3}\\
\bar{C}''+\frac{4\bar{C}'}{r}&=0.\label{C3}
\end{align}
The first and third equations are now linear, and have general
solutions
\begin{align}
\bar{A}&=a_1r+\frac{a_2}{r^2};\label{A3sol}\\
\bar{C}&=\frac{c_1}{r^3}+c_2.\label{C3sol}
\end{align}
One may also substitute eq.~(\ref{A3}) into eq.~(\ref{B3}), such that
the latter simplifies to
\begin{equation}
\frac{4\bar{A}}{r^4}+\frac{2\bar{A}^2}{r^2}+2\bar{C}^2=0.
\label{C3b}
\end{equation}
This constrains the general constants appearing in eqs.~(\ref{A3sol},
\ref{C3sol}), and reproduces precisely the solution of
eq.~(\ref{phisol2}). It seems then that, in contrast to pure
Yang-Mills theory (e.g. ref.~\cite{Singleton:1999ty}), cylindrically
symmetric extended solutions are not possible. \\

All of the solutions presented here are singular as $r\rightarrow0$,
which is not surprising, given that non-singular solutions are
prohibited by Derrick's theorem~\cite{Derrick:1964ww}. There may be
solutions, however, that are regular at the origin, but singular
elsewhere. Such solutions exist in the spherically symmetric case, for
example, as can be seen from the fact that eq.~(\ref{Seq3}) admits the
power series solution
\begin{equation}
S(r)=c-\frac{c^2r^2}{6}+\frac{c^3 r^4}{60}-\frac{11 c^4 r^6}{7560}
+{\cal O}(c^5 r^8),
\label{Spow}
\end{equation}
with $c$ an arbitrary constant. For large enough $r$, however, one may
neglect the second term in eq.~(\ref{Seq3}) so that
\begin{equation}
S''(r)+S^2(r)\simeq 0.
\label{Seq4}
\end{equation}
The solution to this is a Weierstrass elliptic function, which will
have a double pole at some finite value of $r$. Similar solutions
(regular at the origin, singular on a spherical shell at finite $r$)
have been constructed in Yang-Mills theory by
Singleton~\cite{Singleton:1999ty}, where they were compared with the
Schwarzschild solution in gravity.\\

The above analysis suggests that the spectrum of nonperturbative
solutions of biadjoint scalar theory is, perhaps unsurprisingly,
simpler than that of nonabelian gauge theory. This is also true for
scattering amplitudes in both theories, and indeed biadjoint scalar
amplitudes may be used as building blocks for gauge theory
amplitudes~\cite{BjerrumBohr:2012mg}, in order to ensure that the
latter are BCJ dual~\cite{Bern:2008qj} (meaning that the double copy
to gravity is made manifest). If the double and zeroth copies relating
biadjoint, gauge and gravity theories extend beyond perturbation
theory, it is presumably true that nonperturbative solutions in
biadjoint theory can also be used to obtain such solutions in gauge
theory or gravity. To this end, it is useful to compare our solutions
in eqs.~(\ref{phisol1}, \ref{phisol2}) to what appears to be the
closest equivalent in non-abelian gauge theory, namely the Wu-Yang
monopole~\cite{Wu:1967vp}. In a gauge in which $A_0^a=0$, the spatial
components of the gauge field are given by~\cite{Rosen:1972uu}
\begin{equation}
A_i^a=-\frac{\epsilon_{iak}x^k}{er^2}.
\label{Wu-Yang}
\end{equation}
This has cylindrical rather than spherical symmetry and, like
eq.~(\ref{phisol2}), involves an inverse power of the coupling
constant. It is also a pointlike disturbance at the origin, leading to
a divergent field energy there. However, it falls off $\sim r^{-1}$
rather than $\sim r^{-2}$, which behaviour can be traced to the
dimensionality of the coupling constants in the two theories: both
scalar and gauge fields have mass dimension 1, whereas the scalar
coupling $y$ and electromagnetic coupling $e$ have mass dimensions $1$
and $0$ respectively. Thus, there must be an additional power of
$r^{-1}$ in eq.~(\ref{phisol2}) relative to eq.~(\ref{Wu-Yang}) on
dimensional grounds. Despite this minor difference, the forms of
eqs.~(\ref{phisol2}) and eq.~(\ref{Wu-Yang}) are very similar, hinting
at a possible zeroth copy-like relationship that may be underlying
them. This is particularly true for the case $k=2$ in
eq.~(\ref{phisol2}), when it may then be written as
\begin{equation}
\Phi^{aa'}=-\frac{2}{yr^2}\epsilon^{abc}\epsilon^{a'b'c}
\frac{x^b x^{b'}}{r^2}\sim A_i^a A_i^{a'}.
\label{phisol2b}
\end{equation}
That is, the biadjoint solution looks like a product of Wu-Yang gauge
fields, where one traces over the space indices. This may simply be a
coincidence, but in any case the issue of whether there is a zeroth
copy relationship between nonperturbative solutions deserves further
investigation. It is not immediately clear how to find this, given
that in both the amplitude and classical double copies, solutions of
the linearised theory play a crucial role.


\section{Conclusion}
\label{sec:conclude}

Biadjoint scalar theories have arisen in studies of scattering
amplitudes in gauge and gravity theories, such as the CHY
equations~\cite{Cachazo:2013iea,Cachazo:2013hca,Cachazo:2013gna,Cachazo:2013iaa},
and the double copy~\cite{Bern:2008qj,Bern:2010ue,Bern:2010yg}. They
also play an important role in the recently discovered classical
double
copy~\cite{Monteiro:2014cda,Luna:2015paa,Luna:2016due,Ridgway:2015fdl},
which is closely related to the amplitude story. It is
widely hoped that there is a nonperturbative explanation of these
relationships, and to this end it is useful to study nonperturbative
solutions of the biadjoint field equations. The hope is that such
solutions could be used as building blocks for nonperturbative
solutions in gauge and gravity theories, mirroring the role of
biadjoint scalar amplitudes in the perturbative sector.\\

In this paper, we have presented some nontrivial solutions of the
biadjoint field equations, with spherical (eq.~(\ref{phisol1})) and
cylindrical (eq.~(\ref{phisol2})) symmetry. These correspond to
pointlike disturbances localised at the origin. Extended solutions,
with a non-power like form, and also possessing such symmetry, seem not
to be possible. The most closely related gauge theory solution to
those presented here appears to be the Wu-Yang monopole, and the issue
of whether a true zeroth copy-like relationship exists deserves
further investigation. If such a connection can be made, it opens up a
way to understand the nonpeturbative significance of both the zeroth
copy, and the double copy of gauge theories to gravity.

\section*{Acknowledgments}

CDW is supported by the UK Science and Technology Facilities Council
(STFC), under grant ST/L000446/1. He thanks Andr\'{e}s Luna, Donal
O'Connell, Ricardo Monteiro and Isobel Nicholson for ongoing
collaboration on related topics, and comments on the manuscript. He is
also grateful to David Miller and Jack Cribben for useful discussions.

\bibliography{refs.bib}

\providecommand{\href}[2]{#2}\begingroup\raggedright\begin{thebibliography}{10}

\bibitem{Bern:2008qj}
Z.~Bern, J.~Carrasco, and H.~Johansson, ``{New Relations for Gauge-Theory
  Amplitudes},'' {\em Phys.Rev.} {\bf D78} (2008) 085011,
\href{http://www.arXiv.org/abs/0805.3993}{{\tt 0805.3993}}.

\bibitem{Bern:2010ue}
Z.~Bern, J.~J.~M. Carrasco, and H.~Johansson, ``{Perturbative Quantum Gravity
  as a Double Copy of Gauge Theory},'' {\em Phys.Rev.Lett.} {\bf 105} (2010)
  061602, \href{http://www.arXiv.org/abs/1004.0476}{{\tt 1004.0476}}.

\bibitem{Bern:2010yg}
Z.~Bern, T.~Dennen, Y.-t. Huang, and M.~Kiermaier, ``{Gravity as the Square of
  Gauge Theory},'' {\em Phys.Rev.} {\bf D82} (2010) 065003,
  \href{http://www.arXiv.org/abs/1004.0693}{{\tt 1004.0693}}.

\bibitem{Kawai:1985xq}
H.~Kawai, D.~Lewellen, and S.~Tye, ``{A Relation Between Tree Amplitudes of
  Closed and Open Strings},'' {\em Nucl.Phys.} {\bf B269} (1986)
1.

\bibitem{Bern:1998ug}
Z.~Bern, L.~J. Dixon, D.~Dunbar, M.~Perelstein, and J.~Rozowsky, ``{On the
  relationship between Yang-Mills theory and gravity and its implication for
  ultraviolet divergences},'' {\em Nucl.Phys.} {\bf B530} (1998) 401--456,
\href{http://www.arXiv.org/abs/hep-th/9802162}{{\tt hep-th/9802162}}.

\bibitem{Green:1982sw}
M.~B. Green, J.~H. Schwarz, and L.~Brink, ``{N=4 Yang-Mills and N=8
  Supergravity as Limits of String Theories},'' {\em Nucl.Phys.} {\bf B198}
  (1982)
474--492.

\bibitem{Bern:1997nh}
Z.~Bern, J.~Rozowsky, and B.~Yan, ``{Two loop four gluon amplitudes in N=4
  superYang-Mills},'' {\em Phys.Lett.} {\bf B401} (1997) 273--282,
\href{http://www.arXiv.org/abs/hep-ph/9702424}{{\tt hep-ph/9702424}}.

\bibitem{Oxburgh:2012zr}
S.~Oxburgh and C.~White, ``{BCJ duality and the double copy in the soft
  limit},'' {\em JHEP} {\bf 1302} (2013) 127,
\href{http://www.arXiv.org/abs/1210.1110}{{\tt 1210.1110}}.

\bibitem{Carrasco:2011mn}
J.~J. Carrasco and H.~Johansson, ``{Five-Point Amplitudes in N=4
  Super-Yang-Mills Theory and N=8 Supergravity},'' {\em Phys.Rev.} {\bf D85}
  (2012) 025006,
\href{http://www.arXiv.org/abs/1106.4711}{{\tt 1106.4711}}.

\bibitem{Carrasco:2012ca}
J.~J.~M. Carrasco, M.~Chiodaroli, M.~Günaydin, and R.~Roiban, ``{One-loop
  four-point amplitudes in pure and matter-coupled N=4 supergravity},'' {\em
  JHEP} {\bf 1303} (2013) 056,
\href{http://www.arXiv.org/abs/1212.1146}{{\tt 1212.1146}}.

\bibitem{Mafra:2012kh}
C.~R. Mafra and O.~Schlotterer, ``{The Structure of n-Point One-Loop Open
  Superstring Amplitudes},'' {\em JHEP} {\bf 1408} (2014) 099,
\href{http://www.arXiv.org/abs/1203.6215}{{\tt 1203.6215}}.

\bibitem{Boels:2013bi}
R.~H. Boels, R.~S. Isermann, R.~Monteiro, and D.~O'Connell,
  ``{Colour-Kinematics Duality for One-Loop Rational Amplitudes},'' {\em JHEP}
  {\bf 1304} (2013) 107,
\href{http://www.arXiv.org/abs/1301.4165}{{\tt 1301.4165}}.

\bibitem{Bjerrum-Bohr:2013iza}
N.~E.~J. Bjerrum-Bohr, T.~Dennen, R.~Monteiro, and D.~O'Connell, ``{Integrand
  Oxidation and One-Loop Colour-Dual Numerators in N=4 Gauge Theory},'' {\em
  JHEP} {\bf 1307} (2013) 092,
\href{http://www.arXiv.org/abs/1303.2913}{{\tt 1303.2913}}.

\bibitem{Bern:2013yya}
Z.~Bern, S.~Davies, T.~Dennen, Y.-t. Huang, and J.~Nohle, ``{Color-Kinematics
  Duality for Pure Yang-Mills and Gravity at One and Two Loops},''
\href{http://www.arXiv.org/abs/1303.6605}{{\tt 1303.6605}}.

\bibitem{Bern:2013qca}
Z.~Bern, S.~Davies, and T.~Dennen, ``{The Ultraviolet Structure of Half-Maximal
  Supergravity with Matter Multiplets at Two and Three Loops},'' {\em
  Phys.Rev.} {\bf D88} (2013) 065007,
\href{http://www.arXiv.org/abs/1305.4876}{{\tt 1305.4876}}.

\bibitem{Nohle:2013bfa}
J.~Nohle, ``{Color-Kinematics Duality in One-Loop Four-Gluon Amplitudes with
  Matter},''
\href{http://www.arXiv.org/abs/1309.7416}{{\tt 1309.7416}}.

\bibitem{Bern:2013uka}
Z.~Bern, S.~Davies, T.~Dennen, A.~V. Smirnov, and V.~A. Smirnov, ``{Ultraviolet
  Properties of N=4 Supergravity at Four Loops},'' {\em Phys.Rev.Lett.} {\bf
  111} (2013), no.~23, 231302,
\href{http://www.arXiv.org/abs/1309.2498}{{\tt 1309.2498}}.

\bibitem{Naculich:2013xa}
S.~G. Naculich, H.~Nastase, and H.~J. Schnitzer, ``{All-loop infrared-divergent
  behavior of most-subleading-color gauge-theory amplitudes},'' {\em JHEP} {\bf
  1304} (2013) 114,
\href{http://www.arXiv.org/abs/1301.2234}{{\tt 1301.2234}}.

\bibitem{Du:2014uua}
Y.-J. Du, B.~Feng, and C.-H. Fu, ``{Dual-color decompositions at one-loop level
  in Yang-Mills theory},''
\href{http://www.arXiv.org/abs/1402.6805}{{\tt 1402.6805}}.

\bibitem{Mafra:2014gja}
C.~R. Mafra and O.~Schlotterer, ``{Towards one-loop SYM amplitudes from the
  pure spinor BRST cohomology},'' {\em Fortsch.Phys.} {\bf 63} (2015), no.~2,
  105--131,
\href{http://www.arXiv.org/abs/1410.0668}{{\tt 1410.0668}}.

\bibitem{Bern:2014sna}
Z.~Bern, S.~Davies, and T.~Dennen, ``{Enhanced Ultraviolet Cancellations in N =
  5 Supergravity at Four Loop},''
\href{http://www.arXiv.org/abs/1409.3089}{{\tt 1409.3089}}.

\bibitem{Mafra:2015mja}
C.~R. Mafra and O.~Schlotterer, ``{Two-loop five-point amplitudes of super
  Yang-Mills and supergravity in pure spinor superspace},''
\href{http://www.arXiv.org/abs/1505.02746}{{\tt 1505.02746}}.

\bibitem{He:2015wgf}
S.~He, R.~Monteiro, and O.~Schlotterer, ``{String-inspired BCJ numerators for
  one-loop MHV amplitudes},'' {\em JHEP} {\bf 01} (2016) 171,
\href{http://www.arXiv.org/abs/1507.06288}{{\tt 1507.06288}}.

\bibitem{Bern:2015ooa}
Z.~Bern, S.~Davies, and J.~Nohle, ``{Double-Copy Constructions and Unitarity
  Cuts},''
\href{http://www.arXiv.org/abs/1510.03448}{{\tt 1510.03448}}.

\bibitem{Mogull:2015adi}
G.~Mogull and D.~O'Connell, ``{Overcoming Obstacles to Colour-Kinematics
  Duality at Two Loops},'' {\em JHEP} {\bf 12} (2015) 135,
\href{http://www.arXiv.org/abs/1511.06652}{{\tt 1511.06652}}.

\bibitem{Chiodaroli:2015rdg}
M.~Chiodaroli, M.~Gunaydin, H.~Johansson, and R.~Roiban, ``{Spontaneously
  Broken Yang-Mills-Einstein Supergravities as Double Copies},''
\href{http://www.arXiv.org/abs/1511.01740}{{\tt 1511.01740}}.

\bibitem{Saotome:2012vy}
R.~Saotome and R.~Akhoury, ``{Relationship Between Gravity and Gauge Scattering
  in the High Energy Limit},'' {\em JHEP} {\bf 1301} (2013) 123,
\href{http://www.arXiv.org/abs/1210.8111}{{\tt 1210.8111}}.

\bibitem{Vera:2012ds}
A.~Sabio~Vera, E.~Serna~Campillo, and M.~A. Vazquez-Mozo, ``{Color-Kinematics
  Duality and the Regge Limit of Inelastic Amplitudes},'' {\em JHEP} {\bf 1304}
  (2013) 086,
\href{http://www.arXiv.org/abs/1212.5103}{{\tt 1212.5103}}.

\bibitem{Johansson:2013nsa}
H.~Johansson, A.~Sabio~Vera, E.~Serna~Campillo, and M.~Ã. Vázquez-Mozo,
  ``{Color-Kinematics Duality in Multi-Regge Kinematics and Dimensional
  Reduction},'' {\em JHEP} {\bf 1310} (2013) 215,
\href{http://www.arXiv.org/abs/1307.3106}{{\tt 1307.3106}}.

\bibitem{Johansson:2013aca}
H.~Johansson, A.~Sabio~Vera, E.~Serna~Campillo, and M.~A. Vazquez-Mozo,
  ``{Color-kinematics duality and dimensional reduction for graviton emission
  in Regge limit},''
\href{http://www.arXiv.org/abs/1310.1680}{{\tt 1310.1680}}.

\bibitem{Melville:2013qca}
S.~Melville, S.~Naculich, H.~Schnitzer, and C.~White, ``{Wilson line approach
  to gravity in the high energy limit},'' {\em Phys.Rev.} {\bf D89} (2014)
  025009,
\href{http://www.arXiv.org/abs/1306.6019}{{\tt 1306.6019}}.

\bibitem{Monteiro:2011pc}
R.~Monteiro and D.~O'Connell, ``{The Kinematic Algebra From the Self-Dual
  Sector},'' {\em JHEP} {\bf 1107} (2011) 007,
\href{http://www.arXiv.org/abs/1105.2565}{{\tt 1105.2565}}.

\bibitem{Cachazo:2013iea}
F.~Cachazo, S.~He, and E.~Y. Yuan, ``{Scattering of Massless Particles:
  Scalars, Gluons and Gravitons},''
\href{http://www.arXiv.org/abs/1309.0885}{{\tt 1309.0885}}.

\bibitem{Cachazo:2013hca}
F.~Cachazo, S.~He, and E.~Y. Yuan, ``{Scattering of Massless Particles in
  Arbitrary Dimension},''
\href{http://www.arXiv.org/abs/1307.2199}{{\tt 1307.2199}}.

\bibitem{Cachazo:2013gna}
F.~Cachazo, S.~He, and E.~Y. Yuan, ``{Scattering Equations and KLT
  Orthogonality},''
\href{http://www.arXiv.org/abs/1306.6575}{{\tt 1306.6575}}.

\bibitem{Cachazo:2013iaa}
F.~Cachazo, S.~He, and E.~Y. Yuan, ``{Scattering in Three Dimensions from
  Rational Maps},'' {\em JHEP} {\bf 1310} (2013) 141,
\href{http://www.arXiv.org/abs/1306.2962}{{\tt 1306.2962}}.

\bibitem{Mason:2013sva}
L.~Mason and D.~Skinner, ``{Ambitwistor strings and the scattering
  equations},'' {\em JHEP} {\bf 1407} (2014) 048,
\href{http://www.arXiv.org/abs/1311.2564}{{\tt 1311.2564}}.

\bibitem{Geyer:2014fka}
Y.~Geyer, A.~E. Lipstein, and L.~J. Mason, ``{Ambitwistor strings in
  4-dimensions},'' {\em Phys.Rev.Lett.} {\bf 113} (2014) 081602,
\href{http://www.arXiv.org/abs/1404.6219}{{\tt 1404.6219}}.

\bibitem{Casali:2015vta}
E.~Casali, Y.~Geyer, L.~Mason, R.~Monteiro, and K.~A. Roehrig, ``{New
  Ambitwistor String Theories},'' {\em JHEP} {\bf 11} (2015) 038,
\href{http://www.arXiv.org/abs/1506.08771}{{\tt 1506.08771}}.

\bibitem{Geyer:2015bja}
Y.~Geyer, L.~Mason, R.~Monteiro, and P.~Tourkine, ``{Loop Integrands for
  Scattering Amplitudes from the Riemann Sphere},'' {\em Phys. Rev. Lett.} {\bf
  115} (2015), no.~12, 121603,
\href{http://www.arXiv.org/abs/1507.00321}{{\tt 1507.00321}}.

\bibitem{Geyer:2015jch}
Y.~Geyer, L.~Mason, R.~Monteiro, and P.~Tourkine, ``{One-loop amplitudes on the
  Riemann sphere},'' {\em JHEP} {\bf 03} (2016) 114,
\href{http://www.arXiv.org/abs/1511.06315}{{\tt 1511.06315}}.

\bibitem{Monteiro:2014cda}
R.~Monteiro, D.~O'Connell, and C.~D. White, ``{Black holes and the double
  copy},'' {\em JHEP} {\bf 1412} (2014) 056,
\href{http://www.arXiv.org/abs/1410.0239}{{\tt 1410.0239}}.

\bibitem{Luna:2015paa}
A.~Luna, R.~Monteiro, D.~O'Connell, and C.~D. White, ``{The classical double
  copy for Taub–NUT spacetime},'' {\em Phys. Lett.} {\bf B750} (2015)
  272--277,
\href{http://www.arXiv.org/abs/1507.01869}{{\tt 1507.01869}}.

\bibitem{Luna:2016due}
A.~Luna, R.~Monteiro, I.~Nicholson, D.~O'Connell, and C.~D. White, ``{The
  double copy: Bremsstrahlung and accelerating black holes},''
\href{http://www.arXiv.org/abs/1603.05737}{{\tt 1603.05737}}.

\bibitem{Ridgway:2015fdl}
A.~K. Ridgway and M.~B. Wise, ``{Static Spherically Symmetric Kerr-Schild
  Metrics and Implications for the Classical Double Copy},''
\href{http://www.arXiv.org/abs/1512.02243}{{\tt 1512.02243}}.

\bibitem{BjerrumBohr:2012mg}
N.~Bjerrum-Bohr, P.~H. Damgaard, R.~Monteiro, and D.~O'Connell, ``{Algebras for
  Amplitudes},'' {\em JHEP} {\bf 1206} (2012) 061,
\href{http://www.arXiv.org/abs/1203.0944}{{\tt 1203.0944}}.

\bibitem{Rosen:1972uu}
G.~Rosen, ``{Exact solutions to the Yang-Mills field equations},'' {\em J.
  Math. Phys.} {\bf 13} (1972)
595--597.

\bibitem{Prasad:1975kr}
M.~K. Prasad and C.~M. Sommerfield, ``{An Exact Classical Solution for the 't
  Hooft Monopole and the Julia-Zee Dyon},'' {\em Phys. Rev. Lett.} {\bf 35}
  (1975)
760--762.

\bibitem{Bogomolny:1975de}
E.~B. Bogomolny, ``{Stability of Classical Solutions},'' {\em Sov. J. Nucl.
  Phys.} {\bf 24} (1976) 449.
[Yad. Fiz.24,861(1976)].

\bibitem{Julia:1975ff}
B.~Julia and A.~Zee, ``{Poles with Both Magnetic and Electric Charges in
  Nonabelian Gauge Theory},'' {\em Phys. Rev.} {\bf D11} (1975)
2227--2232.

\bibitem{'tHooft:1974qc}
G.~'t~Hooft, ``{Magnetic Monopoles in Unified Gauge Theories},'' {\em Nucl.
  Phys.} {\bf B79} (1974)
276--284.

\bibitem{Polyakov:1974wq}
A.~M. Polyakov, ``{Isomeric States of Quantum Fields},'' {\em Zh. Eksp. Teor.
  Fiz.} {\bf 68} (1975)
1975.

\bibitem{Witten:1976ck}
E.~Witten, ``{Some Exact Multi - Instanton Solutions of Classical Yang-Mills
  Theory},'' {\em Phys. Rev. Lett.} {\bf 38} (1977)
121--124.

\bibitem{Singleton:1999ty}
D.~Singleton, ``{General relativistic analog solutions for Yang-Mills
  theory},'' {\em Theor. Math. Phys.} {\bf 117} (1998) 1351--1363,
\href{http://www.arXiv.org/abs/hep-th/9904125}{{\tt hep-th/9904125}}.

\bibitem{Derrick:1964ww}
G.~H. Derrick, ``{Comments on nonlinear wave equations as models for elementary
  particles},'' {\em J. Math. Phys.} {\bf 5} (1964)
1252--1254.

\bibitem{Wu:1967vp}
T.~T. Wu and C.-N. Yang, ``Some solutions of the classical isotopic gauge field
  equations,'' {\em Properties of Matter Under Unusual Conditions}
(1967).

\end{thebibliography}\endgroup
\end{document}